\documentclass[letterpaper, 10 pt, conference]{ieeeconf}
\IEEEoverridecommandlockouts
\usepackage{caption}
\renewcommand{\thetable}{\arabic{table}}
\usepackage{booktabs}
\usepackage{array}
\usepackage{diagbox}    
\usepackage{cite}
\usepackage{url}
\usepackage{float}
\usepackage{amsmath,amssymb,amsfonts}
\usepackage{algorithmic}
\usepackage{graphicx}
\usepackage{textcomp}
\usepackage{xcolor}
\usepackage{soul}
\usepackage{theorem}
\usepackage{hyperref}
\hypersetup{pdfborder={0 0 0},}
\def\BibTeX{{\rm B\kern-.05em{\sc i\kern-.025em b}\kern-.08em
    T\kern-.1667em\lower.7ex\hbox{E}\kern-.125emX}}
\usepackage{algorithm}
\usepackage{algorithmic}

\newtheorem{theorem}{Theorem}

\begin{document}


\title{\LARGE \bf A Time Splitting Based Optimization Method for Nonlinear MHE
}

\author{Shuting Wu$^\dagger$, Yifei Wang$^\dagger$, Jingzhe Wang, Apostolos I. Rikos and Xu Du$^*$ 
\thanks{$^*$ Corresponding author.}
\thanks{$^\dagger$ The first two authors contributed equally to this work.} 
\thanks{Shuting Wu is with School of Mathematics and Statistics, North China University of Water Resources and Electric Power, Zhengzhou, China. Shuting Wu is also with the HNAS Institute of Mathematics, Henan Academy of Science. E-mail: \texttt{wushuting0126@163.com}.}
\thanks{Yifei Wang is with the Ningbo Artificial Intelligence Institute and the Department of Automation, Shanghai Jiao Tong University, Ningbo 315000, China, and also with the School of Automation and Intelligent Sensing, Shanghai Jiao Tong University, Shanghai 200240, China. E-mail: \texttt{yifeiw4ng@sjtu.edu.cn}.}
\thanks{Jingzhe Wang is with School of Computing and Information, University of Pittsburgh, Pittsburgh, PA, USA. E-mail: \texttt{jiw148@pitt.edu}.}
\thanks{Apostolos I. Rikos and Xu Du are with the Artificial Intelligence Thrust of the Information Hub, The Hong Kong University of Science and Technology (Guangzhou), Guangzhou, China. 
    Apostolos I. Rikos is also affiliated with the Department of Computer Science and Engineering, The Hong Kong University of Science and Technology, Clear Water Bay, Hong Kong, China. E-mails: {\tt~ apostolosr@hkust-gz.edu.cn; michaelxudu@hkust-gz.edu.cn}. } 
 \thanks{The work of A.I.R. and X.D. are also supported by the Guangzhou-HKUST(GZ) Joint Funding Scheme (Grant No. 2025A03J3960).}
}
\maketitle

\begin{abstract}

This paper presents computationally efficient algorithms for solving nonlinear Moving Horizon Estimation (MHE) problems, which face challenges due to the \textit{curse of dimensionality}. Specifically, we first introduce a distributed reformulation utilizing a time-splitting technique. Leveraging this, we develop the Efficient Gauss-Newton Augmented Lagrangian Alternating Direction Inexact Newton (ALADIN) algorithm to improve efficiency. To address limited computational power in some sub-problem solvers, we propose the Efficient Sensitivity Assisted ALADIN, allowing inexact solutions without compromising performance. Additionally, we propose a Distributed Sequential Quadratic Programming (SQP) method for scenarios with no computational resources for sub-problems. Numerical experiments on a differential drive robot MHE problem demonstrate that our algorithms achieve both high accuracy and computational efficiency, meeting real-time requirements.
\end{abstract}



\section{Introduction}
 
Moving Horizon Estimation (MHE) has attracted considerable interest for its applications in differential drive robots \cite{MPC_MHE},
unmanned aerial vehicles \cite{wang2023neural}, and wireless communication \cite{zou2023moving}; a comprehensive overview is provided in \cite{rawlings2017model}. {\color{black}Essentially, MHE is an optimization-based approach for estimating the states of dynamic systems within a moving time horizon}, providing an effective framework for state estimation in nonlinear and constrained dynamic systems. 
Current MHE approaches mainly rely on centralized solvers, yet these methods become computationally prohibitive as estimation complexity and the length of time horizon increase - a challenge commonly described as the \textit{curse of dimensionality}.
{\color{black} To address this challenge, one promising approach is to
reformulate MHE as a distributed optimization problem and adopt parallel algorithms for its solution. However, to the best of our knowledge, a suitable algorithm that efficiently solves distributed MHE has not yet been identified. 
}


A natural approach for solving the distributed optimization reformulation of MHE is to adopt Augmented Lagrangian Alternating Direction Inexact Newton (ALADIN) \cite{Kouzoupis2016}, a distributed non-convex optimization algorithm known for integrating the advantages of Alternating Direction Method of Multipliers (ADMM) \cite{Boyd2011}, \cite{lin2022alternating} and Distributed Sequential Quadratic Programming (SQP) \cite{Nocedal2006}. This motivation arises from ALADIN's demonstrated success in efficiently addressing Model Predictive Control problem~(MPC) \cite{9029760}, \cite{shi2022parallelmpclinearsystems}, \cite{puttschneider2024towards},\cite{stomberg2024decentralized}, \cite{stomberg2024large}- an optimization counterpart of MHE. {\color{black} ALADIN exhibits global convergence for convex problems and local convergence for non-convex problems \cite{Houska2021}, \cite{du2023consensus,Du2025,du2025convergence}, with \cite{du2023bilevel} establishing a global convergence theory for ALADIN in the context of non-convex problems.
}
Typically, ALADIN solves sub-problems using an appropriate nonlinear programming (NLP) solver and coordinates information by solving a coupled quadratic programming~(QP) problem \cite{wang2025aladin}.
However, directly applying standard ALADIN \cite{Houska2016} to MHE remains computationally expensive due to the inherent coupled QP step required for coordinating distributed information, rendering it unsuitable for the real-time requirements of MHE. 
While a variant of ALADIN tailored for MPC\cite{9029760} might be considered, it targets general objective functions (e.g., economic MPC \cite{Diehl2011}) rather than the specific least-squares objective of MHE.
{\color{black}Although a variant of ALADIN, known as Gauss-Newton ALADIN \cite{Du2019}, exists for handling least-squares objectives, it remains computationally inefficient due to the aforementioned} coupled QP step. 
Thus, this gap motivates the following research question: \textit{Can we develop computationally efficient variants of  ALADIN specifically tailored to nonlinear MHE?} 


\subsection*{Contributions}


\noindent\textbf{A.}
In this paper, we introduce a novel time-splitting-based optimization framework for solving nonlinear MHE problems efficiently while maintaining accuracy. We first revisit the nonlinear MHE formulation and propose a time-splitting-based distributed reformulation, extending the temporal decomposition concept originally developed for MPC \cite{9029760}. Our reformulation partitions the time horizon into multiple independent sub-windows, significantly reducing sub-problems dimensionality. 

\noindent\textbf{B.}
Leveraging this distributed reformulation, we develop computationally efficient solutions within the ALADIN framework. Specifically, to eliminate the computational overhead associated with iterative QP solutions required in ALADIN, we first derive a closed-form solution for the QP step. Exploiting this closed-form solution, we propose Efficient Gauss-Newton ALADIN, an accelerated variant of Gauss-Newton ALADIN algorithm introduced in \cite{Du2019}, which achieves computational efficiency. 

\noindent\textbf{C.}
Considering practical scenarios where sub-problem solvers possess limited computational power, we introduce Efficient Sensitivity Assisted ALADIN, inspired by \cite{9993352}, which allows the sub-problems step to be solved inexactly. 

\noindent\textbf{D.}
We further consider an extreme scenario wherein sub-problem solvers have no computational capability. Under this stringent condition, inspired by \cite{stomberg2022decentralized}, we develop an Efficient Distributed SQP that entirely eliminates explicit sub-problem solving. Instead, it only evaluates first- and second-order information of local objectives. 

 \noindent\textbf{E.}
{\color{black}We conducted numerical benchmarks on a practical nonlinear MHE problem involving the differential drive robots.}
The results demonstrate that our Efficient Distributed SQP achieves identical state estimation trajectories to those obtained by \texttt{CasADi}  {\color{black} with \texttt{IPOPT} \cite{MPC_MHE}}. {\color{black} Moreover, all three proposed algorithms exhibit excellent stability in terms of iteration count and convergence precision. Notably, the fastest algorithm achieves high precision in a remarkably short time.} 

\section{Fundamentals of the MHE}
\label{sec: prel-mhe}
\subsection{Discrete Control System}
In control systems, dynamic behavior is typically modeled using discrete-time nonlinear equations, comprising state and output equations that characterize system evolution and observation relationships at time index $n$,
\begin{equation}\label{eq: dynamic system}
\small
    \begin{aligned}
        x_{n+1} &= f(x_n, u_n),\\
        y_n &= h(x_n) + v_n.
    \end{aligned}
\end{equation}
Here, $x_n \in \mathbb{R}^{\lvert x_n \rvert}$ denotes the system state, $u_n \in \mathbb{R}^{\lvert u_n \rvert}$ represents the control input, and $y_n \in \mathbb{R}^{\lvert y_n \rvert}$ stands for the measured output. Note that the measurement noise $v_n$ follows a zero-mean Gaussian distribution, i.e., $v_n \sim \mathcal{N}(0, V)$, where $V$ is a positive-definite covariance matrix. 
Furthermore, the nonlinear dynamics is defined by $f:\mathbb{R}^{|x_n|+|u_n|}\to \mathbb{R}^{|x_n|}$, and the nonlinear measurement function is expressed by $h:\mathbb{R}^{|x_n|}\to \mathbb{R}^{|y_n|}$, both of which are assumed to be twice continuously differentiable.

\subsection{Basics of MHE}
Based on \eqref{eq: dynamic system}, at each time step $l$, given a prediction horizon of length $L$, the following optimization problem represents a formulation of MHE (see \cite{MPC_MHE}): 
\begin{equation}\label{eq: MHE modeling}
\small
    \begin{split}
        &\min_{x, u} \ \frac{1}{2} \left\lVert x_{l-L} - \hat{x}_{l-L} \right\rVert_{P^{-1}}^2\hspace{-0.1cm} + \hspace{-0.1cm}\frac{1}{2} \sum_{n=l-L}^{l} \left\lVert  h(x_n) - y_n \right\rVert_{V^{-1}}^2 \\&\hspace{-0.1cm}+\hspace{-0.1cm}\frac{1}{2}\hspace{-1mm} \sum_{n=l-L}^{l-1}\hspace{-1.2mm} \left\lVert u_n - \hat{u}_n \right\rVert_{W^{-1}}^2
        \hspace{-0.1cm} +\hspace{-0.1cm}\frac{1}{2}\hspace{-1mm} \sum_{n=l-L}^{l-1} \hspace{-1.2mm}\left\lVert x_{n+1} - f(x_n, u_n) \right\rVert_{R^{-1}}^2.
    \end{split}
\end{equation}
The optimization variable is defined as,
\begin{equation}\small\left\{
    \begin{split}
        x &= \left(x_{l-L}^\top, x_{l-L+1}^\top, \dots, x_l^\top\right)^\top,\\
         u &= \left(u_{l-L}^\top, u_{l-L+1}^\top, \dots, u_{l-1}^\top\right)^\top,
    \end{split}\right.
\end{equation}
where $\hat{x}_{l-L}$ represents the prior state estimate {\color{black}\cite[Section 4.2]{rawlings2017model}}, $P\in\mathbb{R}^{|x_{n}|\times |x_{n}|}$ denotes the covariance matrix associated with the initial state estimation error, $R\in\mathbb{R}^{|x_{n}|\times |x_{n}|}$ corresponds to the covariance matrix of the state noise, $V\in\mathbb{R}^{|y_{n}|\times |y_{n}|}$ describes the covariance matrix of the observation noise, and $W\in\mathbb{R}^{|u_{n}|\times |u_{n}|}$ characterizes the covariance matrix of the control input variations. In this expression, the optimization variables of \eqref{eq: MHE modeling} are $x$ and $u$.

An alternative MHE formulation considers only $x$ as the optimization variable. Although $u$ still appears in the expressions, it is treated as a known constant. Based on this, the simplified optimization problem is formulated as follows\footnote{For the convenience of the subsequent expressions, this paper studies MHE without inequality constraints. See \cite{simpson2024parallelizableparametricnonlinearidentification,diehl2009efficient,9993416} for a similar setting.} (see \cite{wynn2014convergence}):
\begin{equation}\label{eq: MHE}\small
\begin{split}
\min_{x}&\frac{1}{2} \left\lVert x_{l-L} - \hat{x}_{l-L} \right\rVert_{P^{-1}}^2\hspace{-0.1cm}+\hspace{-0.1cm}\frac{1}{2} \sum_{n = l-L}^l \left\|h(x_n) - y_n \right\|_{V^{-1}}^2  \\
\text{s.t.} \hspace{0.7mm} & x_{n+1} = f(x_n, u_n),\hspace{0.8cm} \forall n=l-L, \dots, l-1. \\
\end{split}
\end{equation}
This paper focuses on the MHE optimization problem formulated in \eqref{eq: MHE}.


\section{Distributed MHE Reformulation: A Time-Splitting-Based Approach}\label{eq: Reformulation}
This section introduces a time-splitting-based distributed MHE framework built on \eqref{eq: MHE}. By partitioning the time horizon into multiple independent sub-windows, this approach significantly reduces the dimensionality of the sub-problems. 
\subsection{Components of the Time Splitting Reformulation}
\begin{figure*}[h]
	\centering
\includegraphics[width=0.75\textwidth,height=0.125\textheight]{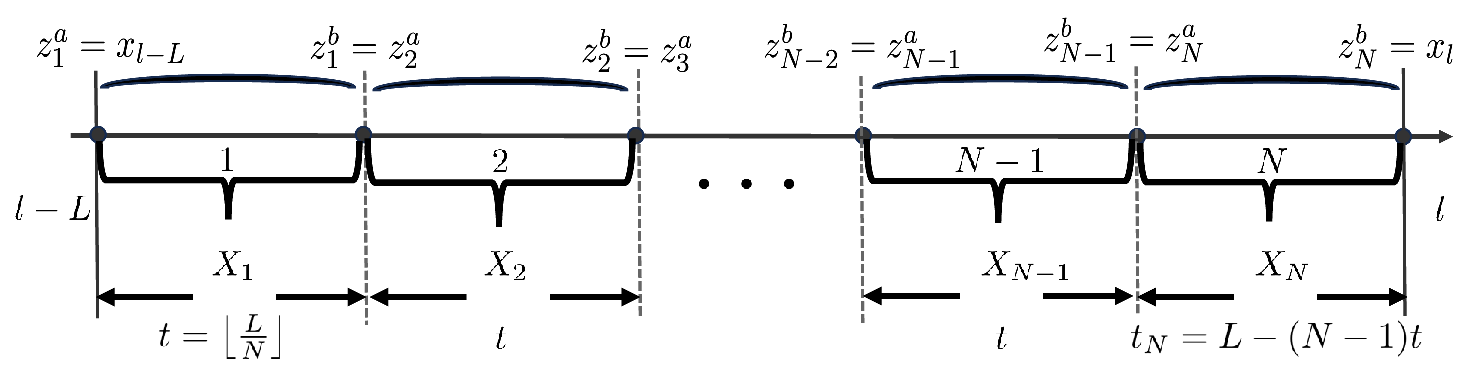}
	\caption{The time-splitting-based MHE}
	\label{fig: time splitting}
\end{figure*}

To mitigate computational complexity and enhance real-time performance in problem \eqref{eq: MHE}, the time window $[l-L, l]$ is divided into $N$ consecutive sub-windows {\color{black}\cite{9029760}}. 
The first $(N-1)$ sub-windows each have a length of $t = \left\lfloor \frac{L}{N} \right\rfloor$, while the last sub-window has a length of $t_N = L - (N-1)t$, where $N, t, t_N \in \mathbb{N}_{>0}$. Accordingly, the time range for the $i$-th sub-window is given by $[l-L+(i-1)t, l-L+it]$, $(i = 1, 2, \dots, N-1)$. For the last sub-window $(i=N)$, the time range is $[l-L+(N-1)t, l]$. Importantly, the auxiliary variable $z = \left((z_1)^\top, (z_2)^\top, \dots, (z_N)^\top\right )^\top$ is introduced to represent the \emph{boundary state} of each sub-window. Here, $z_i = \left((z_i^a)^\top, (z_i^b)^\top\right )^\top$ with $z_i^a$ denoting the \emph{initial state} of the $i$-th sub-window, defined as $z_i^a = x_{l-L+(i-1)t}$. In subsequent sections, $x_{l-L+(i-1)t}$ will be replaced by $z_i^a $. Meanwhile, $z_i^b$ serves as a new auxiliary variable representing the \emph{terminal state} of the $i$-th sub-window.


The optimization variable $X_i$ associated with the local optimization problem for the $i$-th sub-window is defined as: 
\begin{equation*}\small
X_i =\left ((z_i^a)^\top, (\tilde{x}_{(i)})^\top, (z_i^b)^\top\right )^\top, \quad X_i \in \mathbb{R}^{|X_i|},
\end{equation*}
where 
$\tilde{x}_{(i)}$ represents the \emph{internal states} of the $i$-th sub-window, such that $\tilde{x}_{(i)} \in \mathbb{R}^{|\tilde{x}_{(i)}|}$, and is expressed as:
\begin{equation*}\small
\tilde{x}_{(i)}\hspace{-0.15cm} =\hspace{-0.15cm}
\begin{cases}
\left((x_{l-L+(i-1)t+1})^\top\hspace{-0.15cm}, \dots,\hspace{-0.05cm} (x_{l-L+it-1})\hspace{-0.3mm}^\top\right)\hspace{-0.3mm}^\top\hspace{-0.2cm},\hspace{-0.5cm}& i \hspace{-1mm}= \hspace{-1mm}1, \dots, \hspace{-0.5mm}N\hspace{-1mm}-\hspace{-1mm}1, \\
\left((x_{l-L+(N-1)t+1})^\top\hspace{-0.15cm}, \dots,\hspace{-0.05cm} (x_{l-1})^\top\right)^\top\hspace{-0.2cm},\hspace{-1.5cm}& i\hspace{-1mm} =\hspace{-1mm} N.
\end{cases}
\end{equation*}

With the above definitions, a schematic diagram of the \emph{time-splitting-based MHE}, where $z_i^b= z_{i+1}^a$ is illustrated in Figure \ref{fig: time splitting}. Further details are provided in Section \ref{sec: Reformulation of MHE}.

The objective function for each sub-problem is represented by 
$J_i(X_i): \mathbb{R}^{|X_i|} \to \mathbb{R}$, 
and the optimization problem for the $i$-th sub-window is formulated as follows,
for $i = 1$ and $i=N$,
\begin{equation}\label{eq:sub-MHE1}\small
\left  \{\begin{split}
&J_1(X_1)\hspace{-1mm} = \hspace{-1mm}  \frac{1}{2} \| z_1^a \hspace{-1.2mm}-\hspace{-0.7mm} \hat{x}_{l-L} \|_{P^{-1}}^2  
\hspace{-1mm} + \hspace{-1mm}\frac{1}{2}\hspace{-1mm}\sum_{j=l-L}^{l-L+t-1}\hspace{-1.2mm} \|  h(x_j)\hspace{-1mm} - \hspace{-0.7mm}y_j\|_{V^{-1}}^2   ,\\
&J_N(X_N) = \frac{1}{2} \sum_{j = l-L+(N-1)t}^{l} \left\|h(x_j) - y_j \right\|^2_{V^{-1}},
\end{split}\right.
\end{equation}
for $i = 2, \cdots, N-1$,
\begin{equation}\label{eq:sub-MHE2}\small
  \begin{split}
J_i(X_i) = \frac{1}{2} \sum_{j=l-L+(i-1)t}^{l-L+it-1} \|  h(x_j) - y_j\|_{V^{-1}}^2.
\end{split}
\end{equation}
Analogous to the objective function formulation, the nonlinear dynamic equality constraints are partitioned into sub-vectors independently as follows, 
for $i = 1, \cdots, N-1$,
\begin{equation}
 \small
\mathcal{F}_i(X_i)\hspace{-1mm} =\hspace{-1.5mm}
\begin{bmatrix}
x_{l-L+(i-1)t+1}\hspace{-0.5mm} -\hspace{-0.5mm} f(z_i^a, u_{l-L+(i-1)t}) \\
x_{l-L+(i-1)t+2}\hspace{-0.5mm} -\hspace{-0.5mm} f(x_{l-L+(i-1)t+1}, u_{l-L+(i-1)t+1}) \hspace{-0.8mm}\\
\vdots \\
z_i^b\hspace{-0.5mm} -\hspace{-0.5mm} f(x_{l-L+it-1}, u_{l-L+it-1})
\end{bmatrix},
\end{equation}
for $i = N$:
\begin{equation}\label{eq: N-dynalic}
\small
\mathcal{F}_i(X_i)\hspace{-1mm} =\hspace{-1.5mm}
\begin{bmatrix}
x_{l-L+(i-1)t+1}\hspace{-0.5mm} -\hspace{-0.5mm} f(z_i^a, u_{l-L+(i-1)t}) \\
x_{l-L+(i-1)t+2} \hspace{-0.5mm}-\hspace{-0.5mm} f(x_{l-L+(i-1)t+1}, u_{l-L+(i-1)t+1}) \\
\vdots \\
x_l\hspace{-0.5mm} - \hspace{-0.5mm}f(x_{l-1}, u_{l-1})
\end{bmatrix}.
\end{equation}

\subsection{The Time Splitting Reformulation of MHE}\label{sec: Reformulation of MHE}
Consequently, based on \eqref{eq:sub-MHE1}-\eqref{eq: N-dynalic}, the time-splitting-based formulation of MHE can be represented as:
\vspace{-2mm}\begin{equation}\label{eq:reformulation of MHE}\small
\begin{aligned}
\min_{\{X_{i}\}} & \quad \sum_{i=1}^{N} J_i(X_i) \\
\text{s.t.} & \quad \mathcal{F}_i(X_i) = 0 \,\hspace{0.5cm}|\mu_i, \quad \forall i = 1, \cdots, N ,\\
            & \quad \sum_{i=1}^{N} A_i X_i = 0 \,\hspace{0.2cm}|\lambda.
\end{aligned}
\end{equation}
Here, $\mu_i$ represents the dual variable of the sub-constraint $\mathcal{F}_i$, {\color{black}where its dimension is given by,
\vspace{-2mm}\begin{equation*}\small
|\mu_i| =
\begin{cases} 
|X_i|(t-1), & i = 1, 2, \cdots, N-1 ,\\
|X_i|(t_N-1), & i = N,
\end{cases}
\end{equation*}}
while $\lambda\in \mathbb{R}^{(N-1)|z_1^b|}$ denotes the Lagrange multiplier corresponding to the coupling constraints.
The coupling constraint matrix $A_i$ is structurally defined as follows,
\begin{equation*}\small
\begin{split}
A_1& =
\begin{bmatrix}
\bar{\mathbf0} &\hat{\mathbf0} & I_{|z_1^b|} \\
\bar{\mathbf0} &\hat{\mathbf0} &\bar{\mathbf0} \\
\vdots & \vdots & \vdots
\end{bmatrix},
A_N =
\begin{bmatrix}
\vdots & \vdots & \vdots \\
\bar{\mathbf0} & \mathbf0^{(N)} & \bar{\mathbf0} \\
- I_{|z_N^a|} & \mathbf0^{(N)} & \bar{\mathbf0}
\end{bmatrix},\\
\tilde{A}&=
\begin{bmatrix}
- I_{|z_i^a|} & \hat{\mathbf0} & \bar{\mathbf0}\\
\bar{\mathbf0} &\hat{\mathbf0} & \ I_{|z_i^b|} 
\end{bmatrix},  \forall i\in\{2,\cdots N-1\},\\
A_i& =
\begin{bmatrix}
\mathbf0_{|X_i|\times(i-2)|z_1^b|}, &\tilde{A}^\top, & \mathbf0_{|X_i|\times(r-i|z_1^b|)}
\end{bmatrix}^\top,
\end{split}
\end{equation*}
where, matrix $\bar{\mathbf0}=\mathbf0_{|z_1^b|\times|z_1^b|}$; $\hat{\mathbf0}=\mathbf0_{|z_1^b|\times|\tilde{x}_{(1)}|}$; $\mathbf0^{(N)}=\mathbf0_{|z_1^b| \times |\tilde{x}_{(N)}|}$; 
such that $A_1 \in \mathbb{R}^{r \times |X_{1}|}$, $A_i \in \mathbb{R}^{r \times |X_{i}|}$, $A_N \in \mathbb{R}^{r \times |X_{N}|}$, $\tilde{A} \in \mathbb{R}^{2|x_1^b|\times |X_{i}|}$. Note that $\sum_{i=1}^{N} A_i X_i = 0$ contains $z_i^b= z_{i+1}^a$, for $i=1,\cdots,N-1$.

\section{Distributed Optimization Algorithms}
\label{sec: algo}
This section is dedicated to developing efficient solutions within the ALADIN framework to address the time-splitting reformulation of MHE \eqref{eq:reformulation of MHE}. 
Initially, we propose an efficient approach for solving coupled QP, which is integrated into the ALADIN framework. 
Subsequently, based on the aforementioned efficient approach, three ALADIN variants are proposed to reduce the computational burden of the standard ALADIN \cite{Houska2016}. 
In this section, $(\cdot)^+$ denotes the value after the update, whereas $(\cdot)^-$ represents the value before the update.

\subsection{An Efficient Method for Solving Coupled QP}\label{sec: efficient QP}

Before introducing our algorithm for solving problem \eqref{eq:reformulation of MHE}, we first introduce an efficient method for solving the {\color{black}strongly convex QP \eqref{eq: QP} below with coupling constraints:}
\begin{equation}\label{eq: QP}\small
			\begin{split}	\mathop{\mathrm{\min}}_{\{ \Delta X_i\}}\quad& \mathop{\sum}_{i=1}^{N} \frac{1}{2}\Delta X_i^\top H_{i} \Delta X_i+g_{i}^\top \Delta X_i \\
			\mathrm{s.t.} \hspace{1mm}\quad &C_{i} \Delta X_i =0\;
            \hspace{17.5mm}|{\mu_i}, \quad \forall i = 1, \cdots, N ,\\
       &\mathop{\sum}_{i=1}^{N} A_i(X_i^++\Delta X_i)=0\; |{\lambda}	.
			\end{split}
		\end{equation}
\begin{theorem}\label{theorem 1}(Efficient QP)
Let the locally linear independence constraint qualification (LICQ) be satisfied for problem \eqref{eq: QP}, ensuring the linear independence of $C_i$s and $A_i$s for every $i=1, 2,\cdots, N$.
Let the locally second-order sufficient condition (SOSC) \cite{Nocedal2006} be satisfied, i.e., $H_i\succ0, \forall i$. 
Also, let us assume the existence of a unique global optimal solution for problem \eqref{eq: QP}. 
Solving problem \eqref{eq: QP} is equivalent to evaluating the values of $\lambda$, $\mu_i$ and $\Delta X_i$ as follows,  \vspace{-2mm}\begin{equation}\label{eq:the values}\small
			\left\{
			\begin{array}{l}
				\begin{split}
                \lambda =& \left(\mathop{\sum}_{i=1}^{N}G_i - Q_iR_i^{-1}Q_i^\top \right)^{-1}p,\\
                \mu_i= & -R_i^{-1}\left(C_iH_i^{-1}g_i+Q_i^\top \lambda\right),\\
	\Delta X_i= & -H_i^{-1}\left(g_i+C_i^\top \mu_i+A_i^\top \lambda\right),
				\end{split}
			\end{array}
			\right.
		\end{equation} 
 where, \vspace{-3mm}\begin{equation}\small
			\left\{
			\begin{array}{l}
				\begin{split}
                \hspace{-1mm}G_i\hspace{-0.8mm}=&A_iH_i^{-1}A_i^\top, \\
                   \hspace{-1mm}Q_i\hspace{-0.8mm}=&A_iH_i^{-1}C_i^\top, \\
                   \hspace{-1mm}R_i\hspace{-0.8mm}=&C_iH_i^{-1}C_i^\top,
                \end{split}
			\end{array}\right.
\left\{\begin{array}{l}
				\begin{split}
                \hspace{-1mm}q\hspace{-0.8mm}=\hspace{-0.8mm} &\mathop{\sum}_{i=1}^{N}\left(Q_iR_i^{-1}C_i-A_i\right)H_i^{-1}g_i,\\
                   \hspace{-1mm}p\hspace{-0.8mm}=\hspace{-0.8mm}&\mathop{\sum}_{i=1}^{N}A_iX_i^++q.
            \end{split}
			\end{array}
			\right.
		\end{equation}
 
\end{theorem}

\textbf{Proof.} See Appendix \ref{app: Theorem}. \hfill$\blacksquare$

As an extension of Theorem \ref{theorem 1}, we propose the closed-form solution 
\begin{equation}\label{eq:the values2}\small
			\left\{
			\begin{array}{l}
				\begin{split}
                \hspace{-1.5mm}\lambda \hspace{-0.5mm}&=\hspace{-0.8mm} \left(\hspace{-0.5mm}\mathop{\sum}_{i=1}^{N}G_i\hspace{-0.5mm} - \hspace{-0.5mm}Q_iR_i^{-1}Q_i^\top \hspace{-0.5mm}\right)^{-1}\hspace{-1mm}\left(\hspace{-0.5mm}p\hspace{-0.5mm}-\hspace{-0.5mm}\sum_{i=1}^N \hspace{-0.5mm}Q_iR_i^{-1}D_i\hspace{-0.5mm}\right),\\
               \hspace{-1.5mm} \mu_i\hspace{-0.5mm}&= \hspace{-0.8mm} -R_i^{-1}\left(C_iH_i^{-1}g_i+Q_i^\top \lambda-D_i\right),\\
	\hspace{-1.5mm}\Delta X_i\hspace{-0.5mm}&= \hspace{-0.8mm} -H_i^{-1}\left(g_i+C_i^\top \mu_i+A_i^\top \lambda\right).
				\end{split}
			\end{array}
			\right.
		\end{equation} 
of the following problem,
\vspace{-2mm}\begin{equation}\label{eq: QP2}\small
			\begin{split}	\mathop{\mathrm{\min}}_{\{ \Delta X_i\}}\quad& \mathop{\sum}_{i=1}^{N} \frac{1}{2}\Delta X_i^\top H_{i} \Delta X_i+g_{i}^\top \Delta X_i \\
			\mathrm{s.t.} \hspace{1mm}\quad & D_i+ C_{i} \Delta X_i =0\;\hspace{9.5mm}|{\mu_i}, \quad \forall i = 1, \cdots, N ,\\
       &\mathop{\sum}_{i=1}^{N} A_i(X_i^++\Delta X_i)=0\; |{\lambda},
			\end{split}
		\end{equation}
        where $D_i\in \mathbb{R}^{|\mu_i|}$ are given constant matrices. Note that, equations \eqref{eq:the values} and \eqref{eq:the values2} will be integrated into our proposed algorithms.
        Due to space limitations, details are omitted here.

\subsection{Algorithm Development}

Based on \cite{Du2019}, Section~\ref{sec: gaussnewton} introduces an efficient variant of Gauss-Newton ALADIN. Section \ref{sec: Sensitivity Assisted} presents an inexact update version of ALADIN, inspired by \cite{9993352}. 
Finally, Section \ref{sec: Distributed SQP} explores an ALADIN variant in which sub-problems are not locally optimized (drawing inspiration from \cite{stomberg2022decentralized}).

\subsubsection{\textbf{Efficient Gauss-Newton ALADIN}}\label{sec: gaussnewton}

The objective function $J_i(X_i)$ in \eqref{eq:reformulation of MHE} is formulated as a nonlinear least-squares optimization problem, where the full vector-valued measurement function $\mathcal{H}_i(X_i)$ is introduced:
\vspace{-1.5mm}\begin{equation}
\small
\mathcal{H}_i(X_i) = 
\begin{pmatrix} 
P^{-\frac{1}{2}} \left( z_1^a - \hat{x}_{l-L}\right)_{i=1} \\ 
V^{-\frac{1}{2}} \left(h(x_j) - y_j\right)_{j \in \mathcal I_i}
\\V^{-\frac{1}{2}} \left(h(x_l) - y_l\right)_{i=N}\end{pmatrix},
\end{equation}
where, $\mathcal{I}_i = \{l-L+(i-1)t, \cdots,l-L+it-1 \}$. Consequently, the objective function $J_i(X_i)$ of the sub-problems can be expressed as $J_i(X_i) = \frac{1}{2}\|\mathcal{H}_i(X_i)\|^2$.

Efficient Gauss-Newton ALADIN is presented in Algorithm \ref{alg:ALADIN}. 
\begin{algorithm}[h]
	\small
	\caption{Efficient Gauss-Newton ALADIN}
	\textbf{Initialization:} Initial guess of dual variable $\lambda$ and primal variables $\{Y_i\}, \;\forall i$, choose $\rho>0$.\\
    \textbf{Output:} Optimal solution $\{Y_i^\star\}$. \\
	\textbf{Repeat:}
	\begin{enumerate}
		\item Paralleled solve local NLP:
		\begin{equation}\label{ALADIN-step1} \small
        \begin{split}
{X_i}^+\hspace{-1.5mm}=\hspace{-0.5mm}\mathop{\arg\min}_{X_i} &\frac{1}{2}\hspace{-0.5mm}\|\mathcal{H}_i(X_i)\|^2\hspace{-1mm}+\hspace{-0.8mm}\lambda^\top A_iX_i\hspace{-0.8mm}+\hspace{-0.8mm}\frac{\rho}{2}\|X_i\hspace{-1mm}-\hspace{-1mm}Y_i^-\hspace{-0.4mm}\|^2\\
        \mathrm{s.t.} \quad & \mathcal{F}_i(X_i)=0.
        \end{split}
		\end{equation}
        
		\item  Evaluate local variables and sensitivity matrix from $X_i^+$:
         \begin{equation}\label{eq: sensitivity}
         \small
			\left\{
			\begin{array}{l}
				\begin{split}
		b_{i}=&\mathcal{H}_i(X_i^+),\\
B_{i}=&\nabla\mathcal{H}_i(X_i^+)^\top,\\	C_{i}=&\nabla\mathcal{F}_i(X_i^+).
				\end{split}
			\end{array}
			\right.
         \end{equation}
         
         \item Assemble gradient and Hessian:
       \begin{equation} \small
g_i = B_i b_i, \quad H_i = B_i B_i^\top.\end{equation}
		\item Update and broadcast the global dual variable $\lambda$:
		 \begin{equation}\label{eq:global dual variable}\small
                \lambda =\left(\mathop{\sum}_{i=1}^{N}G_i - Q_iR_i^{-1}Q_i^\top \right)^{-1}p.
		\end{equation}  
        
		\item Paralleled update local primal and dual variables:
		\begin{equation}\label{eq:local variables}\small
			\left\{
			\begin{array}{l}
				\begin{split}
					\mu_{i}= & -R_i^{-1}(C_iH_i^{-1}g_i+Q_i^\top \lambda),\\
					Y_{i}^+= &X_i^+ -H_i^{-1}(g_i+C_i^\top \mu_i+A_i^\top \lambda).
				\end{split}
			\end{array}
			\right.
		\end{equation} 
	\end{enumerate}
	\label{alg:ALADIN}
\end{algorithm}
Similar to Gauss-Newton ALADIN \cite{Du2019}, it alternates between solving sub-problems in parallel at the sub-nodes and coordinating via the coupled QP \eqref{eq: QP}. Further, Algorithm \ref{alg:ALADIN} replaces the coupled QP with \eqref{eq:the values}, thereby accelerating computation.
During each iteration, Step $1$) solves the NLP sub-problems \eqref{ALADIN-step1} in parallel using any NLP solver. In Step $2$), each sub-node performs sensitivity analysis based on its local solution, computing the gradient $g_i$ and $H_i$ at each local node according to the optimal solution $X_i^+$, see \eqref{eq: sensitivity}. These results are then transmitted to the central node. After gathering the sensitivity data from all sub-nodes, the central node updates the global dual variable $\lambda$ in Step $4$) using equation \eqref{eq:global dual variable}. The updated $\lambda$ is subsequently broadcast to the sub-nodes, allowing each sub-node to locally update the primal variables according to equation \eqref{eq:local variables}. This process is repeated until convergence.

Note that Algorithm \ref{alg:ALADIN} is specifically tailored for least-squares problems. To extend its applicability and further reduce overall computational time, we propose two additional ALADIN variants designed for broader problem classes.
%


\subsubsection{\textbf{Efficient Sensitivity Assisted ALADIN}}\label{sec: Sensitivity Assisted}

Inspired by \cite{9993352}, we propose Efficient Sensitivity Assisted ALADIN (Algorithm \ref{alg:ALADIN2}) by leveraging the sensitivity of NLP parameters. 

The augmented Lagrangian function for each sub-problem of problem \eqref{eq:reformulation of MHE} is expressed as
\vspace{-2mm}\begin{equation} \label{eq:The augmented Lagrangian}
\small
\begin{split}
\mathcal{L}_i\hspace{-1mm}=\hspace{-1mm}J_i(X_i)\hspace{-0.8mm}+\hspace{-0.8mm}\lambda^\top A_i (X_i\hspace{-0.8mm}-\hspace{-0.8mm}Y_i)\hspace{-0.8mm}+\hspace{-0.8mm}\frac{\rho}{2}\|X_i\hspace{-0.8mm}-\hspace{-0.8mm}Y_i\|^2\hspace{-0.8mm}+\hspace{-0.8mm}\mu_i^\top\mathcal{F}_i(X_i). 
\end{split}
\end{equation}
Following the notation in \cite[IV.C]{9993352}, we define $s_i(\xi_i)=(X_i(\xi_i)^\top,\mu_i(\xi_i)^\top)^\top$ for notational convenience, where $\xi_i=(Y_i^\top,\lambda^\top )^\top$. 
The Karush-Kuhn-Tucker (KKT) conditions for the constrained sub-problems can be further expressed as,
\vspace{-2mm}\begin{equation}\label{eq:KKT conditions}\small
\varphi_i(s_i(\xi_i^-),\xi_i^-) =
\begin{bmatrix}
\nabla_{X_i}\mathcal{L}_i(s_i(\xi_i^-))\\
\mathcal{F}_i(X_i^-)
\end{bmatrix}=0,
\end{equation}
where higher-order terms in the linearization of the solution manifold are neglected, the update for the sub-problems of Algorithm \ref{alg:ALADIN2} is as follows,
\vspace{-2mm}\begin{equation}\label{eq:sensitivity update}\small
s_i^+(\xi_i) = s_i(\xi_i^-) - \mathcal{M}_i^{-1} \mathcal{N}_i\left(\xi_i - \xi_i^-\right),
\end{equation}
where 
$\mathcal{M}_i=
\frac{\partial \varphi_i}{\partial s_i}$, $\mathcal{N}_i=
\frac{\partial \varphi_i}{\partial \xi_i}$. 

Utilizing a tangent predictor, the approximate solutions of the sub-problems at subsequent iterations can be efficiently estimated. Unlike the linearized ALADIN method \cite[Equation (12), Appendix A]{du2023bilevel}, which linearizes the objective function around the current iteration point, this approach instead focuses on linearizing the solution manifold in the vicinity of the parameters.
\vspace{-1mm}\begin{algorithm}[h]
	\caption{Efficient Sensitivity Assisted ALADIN}
	\textbf{Initialization:} Initial guess of dual variable $\lambda$, $\mu_i$, primal variables $\{Y_i=X_i\},\;\forall i$ and parameter $\xi_0=((Y_i)^\top,\lambda^\top)^\top$, choose $\rho>0$.\\
    \textbf{Output:} Optimal solution $\{Y_i^\star\}$.\\
	\textbf{Repeat:}
	\begin{enumerate}
		
	\item Evaluate gradient, Hessian and sensitivity matrix from $X_i$:
       \begin{equation}\label{eq:gradient, Hessian and sensitivity matrix 2}\small
			\left\{
			\begin{array}{l}
				\begin{split}
				g_{i}=&\nabla J_i(X_i),	\\
           H_{i}\approx&\nabla^2 (J_{i}(X_i)+\mu_i^\top \mathcal{F}_i(X_i))+\rho I,\\	C_{i}=&\nabla\mathcal{F}_i(X_i),\\
           D_{i}=&\mathcal F_i(X_i).
				\end{split}
			\end{array}
			\right.
		\end{equation}

		\item  Update and the global dual variable $\lambda$ as \begin{equation}\label{eq: global dual 2}\small
        \small
		    \lambda \hspace{-0.8mm}= \hspace{-0.8mm}\left(\mathop{\sum}_{i=1}^{N}G_i \hspace{-0.6mm}-\hspace{-0.4mm} Q_iR_i^{-1}Q_i^\top \right)^{-1}\hspace{-0.8mm}\left(p\hspace{-0.4mm}-\hspace{-0.8mm}\sum_{i=1}^N Q_iR_i^{-1}D_i\right).
		\end{equation}
        
        \item  Paralleled update $\mu_i$s and $Y_i^+$s
        as \begin{equation}\label{eq:local variables2}\small
			\left\{
			\begin{array}{l}
				\begin{split}
					\hat\mu_{i}= & -R_i^{-1}\left(C_iH_i^{-1}g_i+Q_i^\top \lambda-D_i\right),\\
					Y_{i}^+= &X_i -H_i^{-1}\left(g_i+C_i^\top \hat\mu_i+A_i^\top \lambda\right).
				\end{split}
			\end{array}
			\right.
		\end{equation} 
      \item Collect parameter $\xi_i=\left((Y_i^+)^\top,\lambda^\top \right)^\top$, compute $\mathcal{M}_i,\mathcal{N}_i$ in parallel, and then solve local NLP with \eqref{eq:sensitivity update} \footnotemark{}.
        \item Extract $X_i^+$ from $s_i^+$:
        $s_i^+=((X_i^+)^\top,\mu_i^\top)^\top$.  
	\end{enumerate}
	\label{alg:ALADIN2}
\end{algorithm}

In Algorithm \ref{alg:ALADIN2}, inspired by \eqref{eq:the values2}, the central node updates the global dual variable according to equation \eqref{eq: global dual 2}, incorporating local information from equation \eqref{eq:gradient, Hessian and sensitivity matrix 2}. Each node then concurrently updates its local dual variable $\hat \mu_i$ and primal variable $Y_i^+$ via equation \eqref{eq:local variables2}. Next, each node updates $s_i^+$ using \eqref{eq:sensitivity update}. This process iterates until convergence.

\footnotetext[2]{The update of local primal variables can optionally consist of two phases \cite[Algorithm 1]{9993352}: update using \eqref{eq:sensitivity update} when the KKT condition is almost satisfied; otherwise, update using \eqref{ALADIN-step1}.
}

\subsubsection{\textbf{Efficient Distributed SQP}}\label{sec: Distributed SQP} 

Building on the approach proposed in Decentralized SQP \cite{stomberg2022decentralized}, we propose Efficient Distributed SQP (Algorithm~\ref{alg:ALADIN3}). 
Unlike Algorithm~\ref{alg:ALADIN} and \ref{alg:ALADIN2}, Algorithm \ref{alg:ALADIN3} solves problem \eqref{eq:reformulation of MHE} by bypassing the resolution of sub-problems. 
Moreover, instead of solving the coupled QP \eqref{eq: QP2} via an inner-level ADMM \cite{stomberg2022decentralized}, Algorithm~\ref{alg:ALADIN3} updates the global dual variable $\lambda$, the local variables $\mu_i$ and $\Delta X_i$ according to the closed-form given by \eqref{eq:the values2}.  
\begin{algorithm}[h]
	\caption{Efficient Distributed SQP}
\textbf{Initialization:} Initial guess of dual variable $\lambda$ and primal variables $\{Y_i\},\;\forall i$, choose $\rho>0$.\\
    \textbf{Output:} Optimal solution $\{Y_i^\star\}$.\\
	\textbf{Repeat:}
	\begin{enumerate}
	\item Locally update gradient, Hessian and sensitivity matrix from $Y_i^-$: 
          \begin{equation}\label{eq:gradient, Hessian and sensitivity matrix}\small
			\left\{
			\begin{array}{l}
				\begin{split}
				g_{i}=&\nabla J_i(Y_i^-),	\\
           H_{i}\approx&\nabla^2 (J_{i}(Y_i^-)+\mu_i^\top \mathcal{F}_i(Y_i^-))+\rho I,\\	C_{i}=&\nabla\mathcal{F}_i(Y_i^-),\\
           D_{i}=&\mathcal F_i(Y_i^-).
				\end{split}
			\end{array}
			\right.
		\end{equation}

		\item Update and the global dual as equation \eqref{eq: global dual 2}.
        
         \item  Update $\mu_i$s and $Y_i^+$s
        as \begin{equation}\label{eq:local variables3}\small
			\left\{
			\begin{array}{l}
				\begin{split}
					\mu_{i}= & -R_i^{-1}(C_iH_i^{-1}g_i+Q_i^\top \lambda-D_i),\\
					Y_{i}^+= &Y_i^- -H_i^{-1}(g_i+C_i^\top \mu_i+A_i^\top \lambda).
				\end{split}
			\end{array}
			\right.
		\end{equation} 
	\end{enumerate}
	\label{alg:ALADIN3}
\end{algorithm}

{\color{black}{\subsection{Convergence Analysis}

We now examine the variations and convergence properties of the algorithms presented. 
Specifically, compared to Gauss-Newton ALADIN in \cite{Du2019}, Algorithm~\ref{alg:ALADIN} incorporates the closed-form expression given in \eqref{eq: QP} (as detailed in equations \eqref{eq:global dual variable} and \eqref{eq:local variables}). 
A comprehensive convergence analysis for Gauss-Newton ALADIN is provided in \cite[Theorem 1]{Du2019}.
Algorithm~\ref{alg:ALADIN2} features a local update step inspired from the Sensitivity-Assisted ADMM \cite{krishnamoorthy2020sensitivity}. 
The corresponding convergence analysis will be included in the extended version of this work.
The convergence analysis for Algorithm~\ref{alg:ALADIN3} is derived from \cite[Theorem 1]{stomberg2022decentralized}, and thus, is omitted for brevity.}}

\section{Numerical Experiment}
\label{sec: exp}
In this section, we apply the three proposed algorithms to a practical MHE problem, known as the differential drive robots problem (see \cite{MPC_MHE}). {\color{black}As demonstrated in \cite{Kouzoupis2016}, the MPC problem locally satisfies the conditions of Theorem~\ref{theorem 1}. Given that the MHE problem is shown to be the dual of the MPC problem in \cite[Section 2.2]{Diehl2009c}, it follows that the practical MHE problem also locally satisfies Theorem \ref{theorem 1}.} The following MHE problem involves three state variables, $x=(\phi, \psi, \theta)^\top$, which represent the lateral position $\phi$, longitudinal position $\psi$, and orientation angle $\theta$. Additionally, two control inputs, $u = (v, \omega)^\top$, are considered, where $v$ denotes the linear velocity and $\omega$ the angular velocity. The observation vector $ y=(r,\alpha)^\top$ consists of the relative range $r$ and bearing $\alpha$. Given $x,u,y$ and a sampling time of $T=0.2s$, the dynamics of the MHE system and the observer model are formulated as follows, in contrast to equation \eqref{eq: dynamic system}:
\begin{equation*}\label{eq:robot model dynamics}\small
f(\hspace{-0.4mm}x_n,\hspace{-0.4mm}u_n\hspace{-0.4mm})\hspace{-1mm}=\hspace{-1.4mm}
\begin{bmatrix}
{\phi_n} \\
{\psi_n}\\
{\theta_n}
\end{bmatrix}\hspace{-0.9mm}+\hspace{-0.3mm}T\hspace{-0.8mm}
\begin{bmatrix}
v_n \cos \theta_n \\
v_n\sin \theta_n \\
\omega_n
\end{bmatrix}\hspace{-0.8mm}, y_n\hspace{-1mm}=\hspace{-1.4mm}\begin{bmatrix}
r \\
\alpha
\end{bmatrix}
\hspace{-1.2mm}=\hspace{-1.4mm}
\begin{bmatrix}
\sqrt{\phi_n^2\hspace{-0.8mm}+\hspace{-0.8mm}\psi_n^2} \\
\arctan\hspace{-0.8mm}\left(\hspace{-1mm}\frac{\psi_n}{\phi_n}\hspace{-1mm}\right)
\end{bmatrix}\hspace{-0.8mm}+\hspace{-0.8mm}
\begin{bmatrix}
\nu_r \\
\nu_\alpha
\end{bmatrix}\hspace{-0.8mm},
\end{equation*}
where $\nu_r$ and $\nu_\alpha$ denote Gaussian noise, with $\nu_{r} \sim \mathcal{N}(0, \sigma_{r}^{2})$ and $\nu_{\alpha} \sim \mathcal{N}(0,\sigma_{\alpha}^{2})$. 

\begin{figure}[h]\small
\centering
\includegraphics[width=0.5\textwidth,height=0.28\textheight]{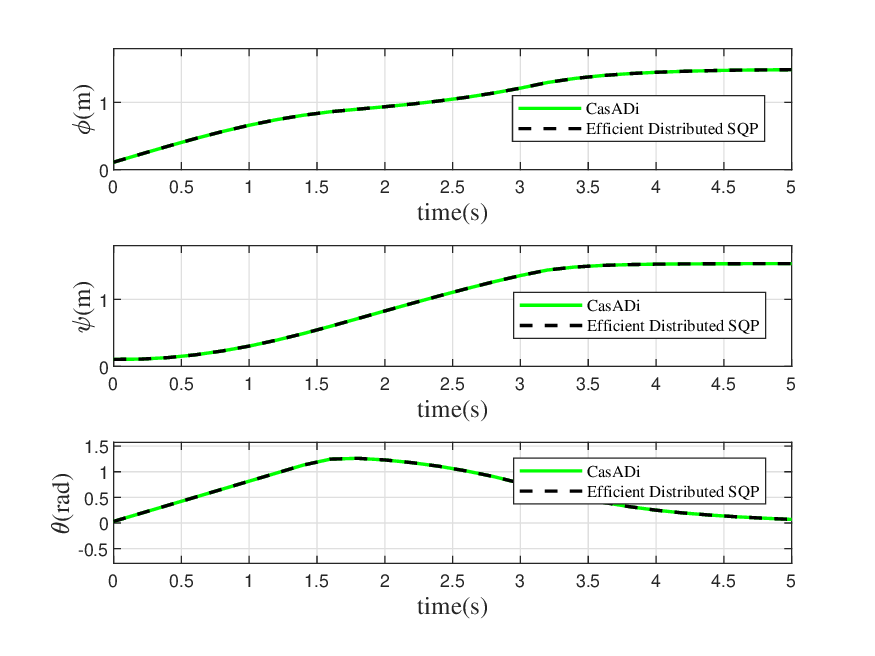}
\caption{Comparison of state estimation trajectories: centralized solver (\texttt{CasADi}) vs. Efficient Distributed SQP.}
\label{fig:state plot}
\end{figure}

The code implementation in this paper is based on \cite{MPC_MHE}. The experimental setup adopts a prediction horizon $L=25$, and the initial states $x_0=(0.1,0.1,0.0)^\top$ define the initial position and orientation of the robot. In the implementation, the state trajectories $x^*=(\phi^*,\psi^*,\theta^*)^\top$ are generated via MPC under the same control model {\color{black}as \cite{MPC_MHE}}. Notably, the primal variables are initialized to $(\phi^*,\psi^*,0)^\top$. In the numerical implementation of Algorithms \ref{alg:ALADIN}, the penalty parameter is set to $\rho=25$, while for Algorithm \ref{alg:ALADIN2} and \ref{alg:ALADIN3}, $\rho=10^3$. The dual variables $\lambda$ and $\mu$ are initialized to zero. All simulations were conducted using \texttt{Casadi-3.6.6} \cite{Andersson2019} with \texttt{IPOPT} in \texttt{MATLAB R2024a} on a Windows $11$ system, equipped with a $2.1$ GHz AMD Ryzen $5$ $4600$U processor and $16$GB of RAM.
Figure \ref{fig:state plot} compares the state trajectories obtained from centralized and distributed solvers for the MHE problem with $N=4$. The results indicate that the proposed distributed MHE framework generates estimates nearly identical to those of the centralized baseline.

\begin{figure}[h]
\centering
\includegraphics[width=0.4\textwidth,height=0.2\textheight]{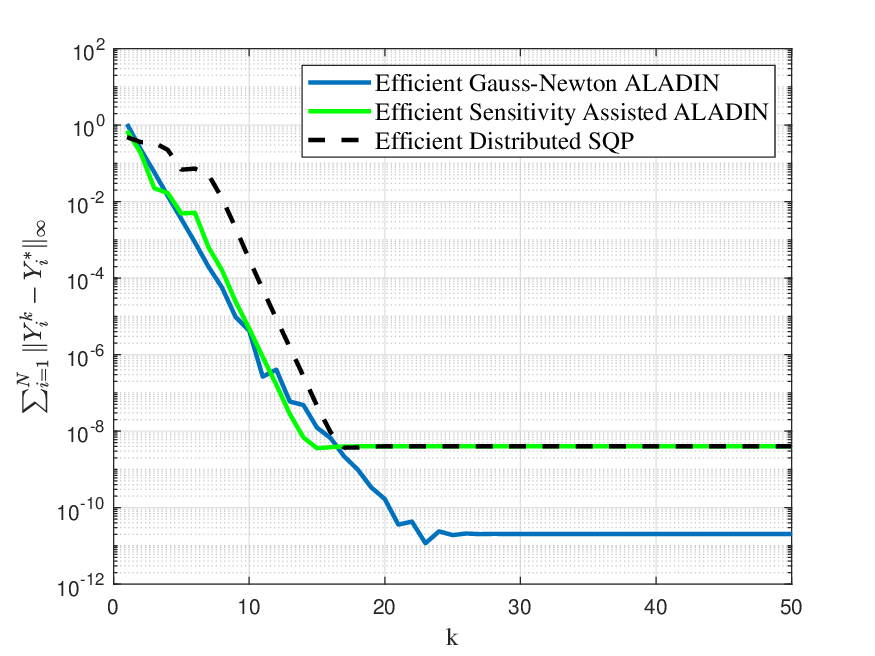}
\caption{Convergence comparison among Algorithms \ref{alg:ALADIN}-\ref{alg:ALADIN3}.}
\label{fig:convergence plot}
\end{figure}

Figure \ref{fig:convergence plot} illustrates the convergence behavior of Algorithm \ref{alg:ALADIN}–\ref{alg:ALADIN3} with $N=4$, all exhibiting linear convergence. Notably, all three algorithms achieve an accuracy of $10^{-8}$ within $30$ iterations, highlighting their computational efficiency. In particular, Algorithm \ref{alg:ALADIN2} leverages \texttt{CasADi} to compute the exact solution in its first iteration, following the recommendations in \cite[Algorithm 1]{9993352}.

Table \ref{tab:CPU time} summarizes the total CPU time of the three proposed efficient ALADIN variants as a function of the number of sub-windows $N$. 
\begin{table}[htbp]
\centering
\small
\renewcommand{\arraystretch}{1}
 \scalebox{0.8}{
\begin{tabular}{ 
  |>{\centering\arraybackslash}m{0.4cm}
  |>{\centering\arraybackslash}m{1.8cm}  
  |>{\centering\arraybackslash}m{1.8cm} 
  |>{\centering\arraybackslash}m{1.8cm}  
  |>{\centering\arraybackslash}m{1.8cm}|
  }
\hline
{\centering\textbf{N}\rule{0pt}{1.2em}}
& \textbf{Algorithm 1} 
& \textbf{Algorithm 2} 
& \textbf{Algorithm 3} 
& \textbf{QP-CasADi} \\
\hline
\textbf{3}\rule{0pt}{1.2em} & 2.83 & 8.51 & 0.0183 & 1.60 \\
\hline
\textbf{4}\rule{0pt}{1.2em} & 2.84 & 8.76 & 0.0184 & 1.71 \\
\hline
\textbf{5}\rule{0pt}{1.2em} & \textbf{2.04} & \textbf{8.08} & \textbf{0.0155} & \textbf{1.57} \\
\hline
\textbf{6}\rule{0pt}{1.2em} & 3.05 & 11.82 & 0.0186 & 1.66 \\
\hline
\end{tabular}
}
\caption{Total CPU time $[s]$ for different algorithms over $N$ sub-windows (measured as the time for $50$ iterations of each algorithm).}
\label{tab:CPU time}
\end{table}
Notably, existing time-splitting-based MPC studies lack theoretical analysis on the relationship between the number of sub-windows $N$ and computational time. By leveraging equations \eqref{eq:the values} and \eqref{eq:the values2}, we establish that the optimal number of sub-windows follows the asymptotic relation $N^*\approx \sqrt{L}$, where $L$ denotes the total horizon length. For brevity, the detailed derivation will be provided in an extended version of this work. In our experiment, setting $L=25$ yields an optimal sub-window count of $N^*=5$. For comparison, we introduce \textbf{QP-CasADi}, which replaces Steps $2$) and $3$) of Algorithm \ref{alg:ALADIN3} with \texttt{CasADi}-based QP solvers. The results demonstrate that across all four algorithmic structures, the configuration with $N=5$ consistently achieves the lowest computational time. {\color{black} The centralized problem \eqref{eq: MHE} was solved in approximately 0.0516 seconds using \texttt{CasADi}}. {\color{black} Notably, although Algorithm \ref{alg:ALADIN2} requires a longer total CPU time than Algorithm \ref{alg:ALADIN}, its sub-problems solutions does not rely on existing solvers, making it particularly suitable for scenarios with limited computational resources at sub-nodes.} 
{\color{black}As we know, \cite{9029760} has already integrated the ALADIN algorithm into the ACADO Toolkit for experimental comparisons in the context of MPC problems. The present study could potentially be extended to conduct similar experiments.
}

\section{Conclusion}
\label{sec: conclusion}
{\color{black}{This paper introduces three computationally efficient distributed optimization algorithms for nonlinear MHE problems, considering sub-problem solver capabilities. We propose a distributed MHE reformulation using a time-splitting strategy and develop new solutions within the ALADIN algorithmic family. By utilizing a closed-form solution for large-scale coupled QP, these algorithms significantly reduce computational time, enabling real-time applications. Numerical experiments on an MHE problem with differential drive robots demonstrate superior convergence and efficiency. Future work will focus on enhancing ALADIN by accelerating matrix updates in Algorithm \ref{alg:ALADIN2} and adaptively prioritizing critical sub-problems, as well as exploring the algorithms' applicability to larger-scale systems.
}}

\appendices
\section{Proof of Theorem \ref{theorem 1}}\label{app: Theorem}
The augmented Lagrangian function for problem \eqref{eq: QP} is defined as: \vspace{-2mm}\begin{equation}\label{eq:The augmented Lagrangian function for the QP problem}
 \small
    \begin{split}
\mathcal{L}(\Delta X_i, \mu_i, \lambda)\hspace{-1mm} =\hspace{-1mm} &\sum_{i=1}^{N} \left (\frac{1}{2} \Delta X_i^\top H_i \Delta X_i + g_i^\top \Delta X_i \right) \\
+&\sum_{i=1}^{N} \mu_i^\top C_i \Delta X_i + \lambda^\top \sum_{i=1}^{N} A_i (X_i^++\Delta X_i ).
\end{split}
    \end{equation} 
From \eqref{eq:The augmented Lagrangian function for the QP problem}, the KKT system of problem \eqref{eq: QP} is given by:
\vspace{-1mm}\begin{equation}\label{eq:KKT conditions for the QP problem}
\small
\left\{
\begin{array}{l}
\frac{\partial \mathcal{L}}{\partial \Delta X_i} = H_i \Delta X_i + g_i + C_i^\top \mu_i + A_i^\top \lambda = 0, \\ 
\frac{\partial \mathcal{L}}{\partial \mu_i} = C_i\Delta X_i = 0, \\ 
\frac{\partial \mathcal{L}}{\partial \lambda} = \mathop{\sum}_{i=1}^{N} A_i (\Delta X_i + X_i^+)= 0.
\end{array}
\right.
\end{equation} 
From the first condition $\frac{\partial \mathcal{L}}{\partial \Delta X_i}=0$ in equation \eqref{eq:KKT conditions for the QP problem}, the following expression is derived:
\vspace{-1mm}\begin{equation}\label{eq:the value of variable} 
\small
\begin{aligned}
 \Delta X_i= -H_i^{-1}(g_i+C_i^\top \mu_i+A_i^\top \lambda)
 \end{aligned}.	
\end{equation} When equation \eqref{eq:the value of variable} is substituted into the second equation of 
 \eqref{eq:KKT conditions for the QP problem}, 
 the resulting equation is expressed as:
\vspace{-1mm}\begin{equation}\label{eq:the value of local variable}\small
\begin{aligned}
\mu_i=-R_i^{-1}(C_iH_i^{-1}g_i+Q_i^\top\lambda) 
\end{aligned}.
\end{equation} Next, by substituting \eqref{eq:the value of local variable} into \eqref{eq:the value of variable} and the third equation of \eqref{eq:KKT conditions for the QP problem}, 
the following result is derived:
\vspace{-2mm}\begin{equation*}
\small
\mathop{\sum}_{i=1}^{N}G_i \lambda=p+\mathop{\sum}_{i=1}^{N}Q_iR_i^{-1}Q_i^\top\lambda.
\end{equation*} 
Through further simplification, the solution for $\lambda$ is obtained as equation \eqref{eq:global dual variable}. 
Subsequently, the local dual variable $\mu_i$ is computed by equation \eqref{eq:the value of local variable} using the previously computed $\lambda$. Finally, the local primal variable increment $\Delta X_i$ is calculated using equation \eqref{eq:the value of variable} based on the obtained $\mu_i$ and $\lambda$. 
Consequently, problem \eqref{eq: QP} has been successfully solved.

\bibliographystyle{IEEEtran}
\bibliography{paper}

\begin{thebibliography}{10}
\providecommand{\url}[1]{#1}
\csname url@samestyle\endcsname
\providecommand{\newblock}{\relax}
\providecommand{\bibinfo}[2]{#2}
\providecommand{\BIBentrySTDinterwordspacing}{\spaceskip=0pt\relax}
\providecommand{\BIBentryALTinterwordstretchfactor}{4}
\providecommand{\BIBentryALTinterwordspacing}{\spaceskip=\fontdimen2\font plus
\BIBentryALTinterwordstretchfactor\fontdimen3\font minus \fontdimen4\font\relax}
\providecommand{\BIBforeignlanguage}[2]{{%
\expandafter\ifx\csname l@#1\endcsname\relax
\typeout{** WARNING: IEEEtran.bst: No hyphenation pattern has been}%
\typeout{** loaded for the language `#1'. Using the pattern for}%
\typeout{** the default language instead.}%
\else
\language=\csname l@#1\endcsname
\fi
#2}}
\providecommand{\BIBdecl}{\relax}
\BIBdecl

\bibitem{MPC_MHE}
M.~G. Mehrez, ``{MPC} and {MHE} implementation in {Matlab} using {CasADi},'' \emph{github}, 2022.

\bibitem{wang2023neural}
B.~Wang, Z.~Ma, S.~Lai, and L.~Zhao, ``Neural moving horizon estimation for robust flight control,'' \emph{IEEE Transactions on Robotics}, vol.~40, pp. 639--659, 2023.

\bibitem{zou2023moving}
L.~Zou, Z.~Wang, B.~Shen, and H.~Dong, ``Moving horizon estimation over relay channels: Dealing with packet losses,'' \emph{Automatica}, vol. 155, p. 111079, 2023.

\bibitem{rawlings2017model}
J.~B. Rawlings, D.~Q. Mayne, M.~Diehl \emph{et~al.}, \emph{Model predictive control: theory, computation, and design}.\hskip 1em plus 0.5em minus 0.4em\relax Nob Hill Publishing Madison, WI, 2017, vol.~2.

\bibitem{Kouzoupis2016}
D.~Kouzoupis, R.~Quirynen, B.~Houska, and M.~Diehl, ``A block based {ALADIN} scheme for highly parallelizable direct optimal control,'' in \emph{American Control Conference}, 2016, pp. 1124--1129.

\bibitem{Boyd2011}
S.~Boyd, N.~Parikh, E.~Chu, B.~Peleato, and J.~Eckstein, ``Distributed optimization and statistical learning via the alternating direction method of multipliers,'' \emph{Found. Trends Mach. Learn.}, vol.~3, no.~1, pp. 1--122, 2011.

\bibitem{lin2022alternating}
Z.~Lin, H.~Li, and C.~Fang, \emph{Alternating direction method of multipliers for machine learning}.\hskip 1em plus 0.5em minus 0.4em\relax Springer, 2022.

\bibitem{Nocedal2006}
J.~Nocedal and S.~Wright, \emph{Numerical optimization}.\hskip 1em plus 0.5em minus 0.4em\relax Springer Science \& Business Media, New York, 2006.

\bibitem{9029760}
Y.~Jiang, C.~N. Jones, and B.~Houska, ``A time splitting based real-time iteration scheme for nonlinear {MPC},'' in \emph{IEEE Conference on Decision and Control}, 2019, pp. 2350--2355.

\bibitem{shi2022parallelmpclinearsystems}
\BIBentryALTinterwordspacing
J.~Shi, Y.~Jiang, J.~Oravec, and B.~Houska, ``Parallel {MPC} for linear systems with state and input constraints,'' 2022. [Online]. Available: \url{https://arxiv.org/abs/2207.00037}
\BIBentrySTDinterwordspacing

\bibitem{puttschneider2024towards}
J.~P{\"u}ttschneider, J.~Golembiewski, N.~A. Wagner, C.~Wietfeld, and T.~Faulwasser, ``Towards event-triggered {NMPC} for efficient {6G} communications: Experimental results and open problems,'' \emph{arXiv preprint arXiv:2409.18589}, 2024.

\bibitem{stomberg2024decentralized}
G.~Stomberg, A.~Engelmann, M.~Diehl, and T.~Faulwasser, ``Decentralized real-time iterations for distributed nonlinear model predictive control,'' \emph{arXiv preprint arXiv:2401.14898}, 2024.

\bibitem{stomberg2024large}
G.~Stomberg, M.~Raetsch, A.~Engelmann, and T.~Faulwasser, ``Large problems are not necessarily hard: A case study on distributed {NMPC} paying off,'' \emph{arXiv preprint arXiv:2411.05627}, 2024.

\bibitem{Houska2021}
B.~Houska and Y.~Jiang.\hskip 1em plus 0.5em minus 0.4em\relax Springer, 2021, ch. Distributed Optimization and Control with {ALADIN}, pp. 135--163.

\bibitem{du2023consensus}
X.~Du and J.~Wang, ``Consensus {ALADIN}: A framework for distributed optimization and its application in federated learning,'' \emph{arXiv preprint arXiv:2306.05662}, 2023.

\bibitem{Du2025}
------, ``Distributed consensus optimization with consensus {ALADIN},'' in \emph{American Control Conference}, 2025 (accepted for publication).

\bibitem{du2025convergence}
X.~Du, X.~Zhou, S.~Zhu, and A.~I. Rikos, ``Convergence theory of flexible {ALADIN} for distributed optimization,'' in \emph{European Control Conference (ECC)}, 2025 (accepted for publication).

\bibitem{du2023bilevel}
X.~Du, J.~Wang, X.~Zhou, and Y.~Mao, ``A bi-level globalization strategy for non-convex consensus {ADMM} and {ALADIN},'' 2023.

\bibitem{wang2025aladin}
Y.~Wang, S.~Wu, G.~Yang, J.~Chu, A.~I. Rikos, and X.~Du, ``Aladin-$\beta$: A distributed optimization algorithm for solving {MPCC} problems,'' in \emph{Conference on Decision and Control (CDC)}, 2025 (accepted for publication).

\bibitem{Houska2016}
B.~Houska, J.~Frasch, and M.~Diehl, ``An augmented {L}agrangian based algorithm for distributed {nonconvex} optimization,'' \emph{SIAM Journal on Optimization}, vol.~26, no.~2, pp. 1101--1127, 2016.

\bibitem{Diehl2011}
\BIBentryALTinterwordspacing
M.~Diehl, R.~Amrit, and J.~B. Rawlings, ``A {L}yapunov function for economic optimizing model predictive control,'' \emph{IEEE Transactions on Automatic Control}, vol.~56, no.~3, pp. 703--707, March 2011. [Online]. Available: \url{http://ieeexplore.ieee.org/xpls/abs\_all.jsp?arnumber=5672577}
\BIBentrySTDinterwordspacing

\bibitem{Du2019}
X.~Du, A.~Engelmann, Y.~Jiang, T.~Faulwasser, and B.~Houska, ``Distributed state estimation for {AC} power systems using {G}auss-{N}ewton {ALADIN},'' in \emph{IEEE Conference on Decision and Control}, 2019, pp. 1919--1924.

\bibitem{9993352}
D.~Krishnamoorthy and V.~Kungurtsev, ``A sensitivity assisted alternating directions method of multipliers for distributed optimization,'' in \emph{IEEE Conference on Decision and Control}, 2022, pp. 295--300.

\bibitem{stomberg2022decentralized}
G.~Stomberg, A.~Engelmann, and T.~Faulwasser, ``Decentralized non-convex optimization via bi-level {SQP} and {ADMM},'' in \emph{IEEE Conference on Decision and Control}, 2022, pp. 273--278.

\bibitem{simpson2024parallelizableparametricnonlinearidentification}
\BIBentryALTinterwordspacing
L.~Simpson, J.~Asprion, S.~Muntwiler, J.~Köhler, and M.~Diehl, ``Parallelizable parametric nonlinear system identification via tuning of a moving horizon state estimator,'' 2024. [Online]. Available: \url{https://arxiv.org/abs/2403.17858}
\BIBentrySTDinterwordspacing

\bibitem{diehl2009efficient}
M.~Diehl, H.~J. Ferreau, and N.~Haverbeke, ``Efficient numerical methods for nonlinear mpc and moving horizon estimation,'' \emph{Nonlinear model predictive control: towards new challenging applications}, pp. 391--417, 2009.

\bibitem{9993416}
K.~Baumgärtner, R.~Reiter, and M.~Diehl, ``Moving horizon estimation with adaptive regularization for ill-posed state and parameter estimation problems,'' in \emph{IEEE Conference on Decision and Control}, 2022, pp. 2165--2171.

\bibitem{wynn2014convergence}
A.~Wynn, M.~Vukov, and M.~Diehl, ``Convergence guarantees for moving horizon estimation based on the real-time iteration scheme,'' \emph{IEEE Transactions on Automatic Control}, vol.~59, no.~8, pp. 2215--2221, 2014.

\bibitem{krishnamoorthy2020sensitivity}
D.~Krishnamoorthy and V.~Kungurtsev, ``Sensitivity assisted alternating directions method of multipliers for distributed optimization and statistical learning,'' \emph{arXiv preprint arXiv:2009.05845}, 2020.

\bibitem{Diehl2009c}
M.~Diehl, H.~J. Ferreau, and N.~Haverbeke, ``Efficient numerical methods for nonlinear {MPC} and moving horizon estimation,'' in \emph{Nonlinear model predictive control}, ser. Lecture Notes in Control and Information Sciences, L.~Magni, M.~Raimondo, and F.~Allg\"ower, Eds.\hskip 1em plus 0.5em minus 0.4em\relax Springer, 2009, vol. 384, pp. 391--417.

\bibitem{Andersson2019}
J.~A.~E. Andersson, J.~Gillis, , G.~Horn, J.~B. Rawlings, and M.~Diehl, ``{CasADi}: a software framework for nonlinear optimization and optimal control,'' \emph{Mathematical Programming Computation}, vol.~11, no.~1, pp. 1--36, Mar 2019.

\end{thebibliography}
		
\end{document}